# The level of non-thermal velocity fluctuations deduced from Doppler spectroscopy and its role on TJ-II confinement

B. Zurro, A. Baciero and TJ-II Team
Laboratorio Nacional de Fusión. Asociación Euratom/Ciemat para Fusión
E-28040 Madrid. Spain
b.zurro@ciemat.es

## Abstract

The goal of this investigation is to study, in the line of previous works, the level of velocity fluctuations in different scenarios of the TJ-II stellarator. The method followed consists in measuring the apparent Doppler temperature of $C^{4+}$ and protons with high spectral resolution techniques with spatial resolution. The level of turbulent velocities in the plasma has been deduced from the difference observed between the apparent temperature of both species, following a method previously presented and borrowed from astrophysics. The study of this difference, as a function of plasma density and injected power, provides a way to explore if this turbulence plays any role in the confinement of the hot TJ-II plasma.

## 1 Introduction

The confinement scaling laws in stellarators depict general trends [1], which need to be understood from basic mechanism accounting for them. Global confinement time studies in TJ-II permits its parameterization in the same way as data from other devices, but as anywhere, its behavior with density, absorbed power etc., is difficult to understand from the point of view of neoclassical or any other theory. In addition, we do not dispose of a simple parameter or small set of them that can be used to predict the global confinement clearly. The study of velocity fluctuation levels in the core of TJ-II intends to contribute to shed some light on these mechanisms and to figure out whether they follow any trend comparable to that of global confinement.

These velocity fluctuations, or nonthermal velocities as termed in astrophysics, to which the spectral line width are sensible with a different strength depending on the ion mass, can be difficult to observe using standard techniques to monitor plasma fluctuations, since they are typically sensitive mainly to density fluctuations. Therefore, the limitations of the present approach are compensated for its uniqueness and by the use of high-resolution spectral techniques available for other purposes.

The goal of this investigation is to study, in the line of previous works [2-4], the level of velocity fluctuations in different scenarios of the TJ-II stellarator. The method followed consists in measuring with spatial resolution the Doppler broadening of protons and $C^{4+}$ with spatial resolution, such as it has been explained in former publications [5-7]. The level of turbulent velocities in the plasma has been deduced from the difference observed between the apparent temperature of both species, following the method previously presented in [2] and borrowed from astrophysics. The study of this difference, as a function of plasma density and injected power, provides a way to explore whether this turbulence plays any role or at least correlates with the confinement trend of the hot TJ-II plasma.

## 2 Experimental

This experiment has been carried out in the TJ-II flexible heliac. It is a four-period, low magnetic shear stellarator with major and averaged minor radii of 1.5 m and $\leq$ 0.22 m, respectively. Central electron densities and temperatures up to $1.7 \times 10^{19}$ $m^{-3}$ and 1 keV respectively are achieved for plasmas created and maintained by ECRH at the second harmonic ($f$ = 53.2 GHz, $P_{ECRH} \leq$ 400 kW). Additional heating is provided by two neutral beam injectors: NBI_1 and NBI_2 parallel and anti-parallel to the toroidal field (up to 400 kW from each NBI). With NBI heating, the typical central density and electron temperature are 2-6x$10^{19}$ $m^{-3}$ and 300-400 eV.

The spectral line shapes of intrinsically impurity ions have been recorded in TJ-II plasmas by means of high spectral resolution spectrometers viewing the plasma perpendicular to the main magnetic field by means of fiber guides. The spatial resolved measurements are achieved with 9 equal-spaced channels separated 25 mm and with a horizontal orientation as shown in Fig. 1 of reference [5]. The experimental system and data analysis methods are described in [6] and [7].

The main assumptions implicit in Doppler broadening studies of ion temperature are: impurity ions are thermalized with protons: the Zeeman effect is negligible or can be corrected for; and other broadening mechanisms, such as Stark effect, are negligible because of the low densities involved in this magnetic confinement device. The thermalization time between protons and other ions has been estimated assuming solely Coulomb collisions [8] and is fast enough to keep the same kinetic temperature for all the ions: impurity and protons.

## 3 The nonthermal velocities fluctuations and its differential influence on spectral line widths

The idea that plasma turbulence can influence spectral line width was suggested many years ago by Struve. Since then, it has been widely used, and with increasing sophistication, in astrophysical plasmas not only to understand the Doppler line width of emission and absorption lines, but also to quantify the level of velocity fluctuations caused by waves and stochastic turbulence in a variety of stellar objects. In particular, many papers can be found on this subject with the goal of understanding the anomalous coronal heating; as a recent reference on this problem we can mention, for instance [9] and references therein. In contrast, less work has been carried out on this subject in laboratory plasmas, although many experimental observations are very difficult to understand if you do not consider the nonthermal velocities effect on Doppler broadening; in particular, in ranges of temperatures where they can be observable. We have dedicated some effort to this problem previously [2-4] and now that we have spectral system with spatial resolution we have started to look at the TJ-II spectroscopic data from this perspective. The first attempt presented here tries to show the clearest evidence of its influence and its correlation with interesting aspects of TJ-II confinement. In a rather low density ($\leq 5x10^{19}$ $m^{-3}$) and magnetic field ($\leq$ 1 T) device, where the dominant emission line broadening mechanism is the Doppler effect and where the main ions and impurities should be thermalized because of its strong collisional coupling, the observation of different temperatures of an impurity ion and protons is the best indicator that the turbulence effect must be present. Effect of plasma turbulence on spectral line width, was considered, within the field of magnetic confined plasmas by Fussmann [10], who worked out a handy formula to apply when working with experimental data. The apparent temperature of an impurity ion of mass $m_z$, $T_z$, can be written as a function of the true kinetic ion temperature of a proton plasma ($m_p$ is the proton mass), $T_p$, and the turbulence temperature, $T_{turb}$, as:

$$T_z = T_p + m_z/m_p \, T_{turb} \qquad (1)$$

Therefore, a "turbulence temperature" of 10 eV (astrophysicists prefer to talk in terms of nonthermal velocities), would produce for a highly ionized carbon present in the core of a high temperature plasma an apparent temperature, if you do not consider this effect, that is 120 eV higher than that one should expect if the turbulence velocities would not be present. Notice the effect of the ion mass on the measured impurity temperature. Of course, this effect would be more difficult to observe if we are dealing with temperatures of a few keV's. However, in the later case the effect would be observable if one uses a Fe line for which the effect would be 550 eV for the same level of velocity fluctuations. Several years ago, Von Hellermann [11], using a pair of values of temperatures of several ions with different masses, obtained values of $T_{turb}$, although he interpreted them as temperature fluctuations and not as nonthermal velocity fluctuations in the astrophysical sense. Their values were similar to those reported in this work.

## 4 Comparisons between global behavior of C V and proton temperatures

In order to illustrate the effect discussed here, we will present first central chord data displayed as a function of density. Afterwards we will focus on a few selected profiles of the type we are going to use to show the structure of the turbulence and its correlation with plasma confinement. We depict, in Fig. 1 (a), as a function of the line-averaged density the behavior of the $C^{4+}$ temperature along a central chord deduced from its Doppler broadening, ignoring for the moment the possible effect of plasma turbulence; their values correspond to the full points of this last figure. The proton temperature obtained, by an independent and well proved traditional diagnostic such as a neutral particle analyzer (NPA), also along a central TJ-II chord, is represented by the full triangles in the same Fig. 1 (a). Specific details regarding to the latter diagnostic and its results can be found in [12] and references therein. Even if we do not have common points at all densities, the trend behavior of those data sets is clear in particular at densities beyond $1x10^{19}$ $m^{-3}$, where we will focus, in the posterior part of this work, the physical discussion. The lower electron temperatures ($T_e(0) \leq 400$ eV) in this density range ($\leq 400$ eV) favors that the $C^{4+}$ emission is coming from the core although if doing the correction due to the chord averaging effect is almost constant and a factor 1.4 higher; what goes in the direction that the observed effect in chord averaged results would be higher with local values. On the other hand, since the NPA data analysis is carried out, after doing the appropriated corrections, in an energy range a factor 5 to 10 higher than the expected ion temperature, they represent directly local values.

A similar effect is seen if we do a similar plot with temperature data of $C^{4+}$ and protons, Fig. 1 (b), obtained in this case directly from spectroscopy in the way described in reference [7]. The proton temperatures obtained by the spectroscopic method are very similar to those obtained by the NPA, as can be seen in the data depicted in Fig. 1 (c), where we have plotted spectroscopic deduced values versus NPA temperature results for a set of TJ-II discharges. The spectroscopic data in that figure are, on averaged, approximately 10 % higher than those provided by the NPA. Since the spatial analysis of the fast neutral emission, from which the spectroscopic data are deduced, are much more difficult to handle than the impurity emission, we have used this heuristic method to justify the validity of the spectral method to study the proton temperature profile in the particular conditions of TJ-II plasmas and also to justify that it provides local values equivalent to those delivered by the NPA. It is convenient to remark that the $C^{4+}$ temperatures, depicted in Fig. 1 (b), agree quite well with CXRS local temperatures of $C^{6+}$ obtained in the same experimental campaign [13].

The difference between $C^{4+}$ and proton temperatures is not exhibited in some sequences of discharges, such as those included in the plots of Fig. 2. We observe in that sequence that for some discharges the $C^{4+}$ temperatures fall within the trend shown by the proton temperature data and that is true for several densities, as shown in the plot of Fig. 2 (a). If we perform the plot of the same points as function of the soft X ray central-chord value, which represent a continuous indication of the discharge temperature, we can see in Fig. 2 (b) that the $C^{4+}$ temperature points which fall within the trend of the proton temperature correspond at discharge times where the electron temperature is lower. This observation that we have seen in more cases, in particular when the discharge shortens and the single shot trigger of the spectrometer CCD was not changed, indicates that significant and easily observable velocity fluctuations capable of producing the reported effect are an intrinsic property of high temperature discharges of TJ-II, that may be not present below some threshold of the electron temperature. Consequently, the density is not the only key parameter to depict this effect.

## 5 Correlation with confinement features

In order to obtain spatially resolved information on turbulent velocities in the core of the TJ-II device, we have focus our study of the nonthermal velocities on NBI discharges with line-averaged densities beyond $1x10^{19}$ $m^{-3}$. There are several reasons to choose this density range for the present purpose. First, the spectral signal available is much higher than in the ECRH phase, in particular with the lithium wall presently used in TJ-II, and therefore the statistical error bars are very small, what favors the use of equation (1) to deduce the "turbulence temperature". Second, the TJ-II central electron temperature in that density range is lower than in the ECRH phase, and ranging between 400 and 250 eV, what is very convenient since the $C^{4+}$ ion emits from the plasma interior and it is consequently a good probe of this effect in the core, whereas for its application in the ECRH phase would be indispensable to invert the chord averaged results in order to obtain local values, increasing the error propagation in the determination of the turbulence level through the formula (1).

We present in Fig. 3 (a) and (b) typical profiles of $C^{4+}$ and proton temperatures obtained by the spectral method, for two selected plasma densities. From the application of formula (1) to these data, we obtain the "turbulence temperature" profiles depicted in Fig. 3 (c) and (d). By using turbulence profiles of this type and taking an averaged of the seven channels we have obtained the plot of Fig. 4 (a) where we can see the density evolution of the averaged values as a function of density for a set of discharges. Superimposed to the discrete points of the former plot we have shown the typical dependence of total energy provided by the diamagnetic loop. This plot shows that there is a correlation between the decrease of the turbulence and the rise in the total energy when increasing the line-averaged electron density.

A paramount signature of global energy confinement in magnetic fusion devices is its degradation with the absorbed power, reflected in the scaling law ($\tau_E$ α $P^{-0.61}$) from the stellarators data base [1]. In order to evaluate whether the nonthermal velocity fluctuations correlate with that degradation of confinement, we have chosen similar discharges where temperature profiles of $C^{4+}$ and protons were taken in similar conditions, in one case with two NBI and in the other with only one injector. The ion temperature profiles for these two discharges, as well as the $T_{turb}$ profiles deduced from them, are plotted in Figs 5 (a), (b) and (c), respectively. The ratio of the averaged values of $T_{turb}$ in those two cases is 0.43 and for the nonthermal velocities ($\alpha\ T^{1/2}$) that ratio is 0.66, which is roughly equal to $2^{-0.61} = 0.65$, assuming that the absorption and power of both NBI are similar.

In conclusion, we have shown how the role of nonthermal velocity fluctuations is important for understanding the behavior of impurity and proton temperatures in the typical conditions of TJ-II plasmas. In addition, we have illustrated how the nonthermal velocities can be correlated with confinement parameters highlighting its potential importance for the core physics of TJ-II plasmas. We expect to perform, in the future, a more detailed study following the line outlined in the present work.

**Acknowledgements.** This work was partially funded by the Spanish "Ministerio de Educación y Ciencia" under Grant No. ENE2007-65007. Drs. J. M. Fontdecaba, E. Ascasíbar and F. Medina are gratefully acknowledged for NPA, soft X-Ray and diamagnetic energy data, respectively.

**Figure Captions**

**Figure 1.** Comparison plots of central $C^{4+}$ temperature and proton temperatures measured by either the NPA or the spectroscopic method: a) plot of $C^{4+}$ temperature NPA temperature as a function of the line-averaged density for a set of TJ-II discharges belonging to the 2008 experimental campaign; b) Similar plot but for a set of discharges of the 2009 campaign; the proton temperature deduced from the spectroscopic method is plotted here; c) Plot showing a comparison of central proton temperatures obtained by the two independent methods used in this work: NPA and spectroscopic methods.

**Figure 2.** Temperature data in a sequence where for some discharges there are good agreement between $C^{4+}$ and proton temperature: a) Plot of their temperatures as a function of $n_e$; b) Similar plot, but as a function of the central soft-X ray monitor signal as a continuous monitor of the plasma temperature.

**Figure 3.** a) and b) Plots of impurity and proton temperatures profiles at two different densities and c) and d) The deduced profiles of turbulent temperature, according to equation (1), for those two densities.

**Figure 4.** Correlation between the averaged turbulent levels deduced at different densities and the behavior of the total diamagnetic energy content in the same sequence of discharges.

**Figure 5.** Comparison of temperature profiles ($C^{4+}$ and protons) and deduced turbulence profiles in NBI discharges, for two similar densities: a) Two injectors b) One injector and c) The deduced turbulence profiles for the two cases.

Fig. 1

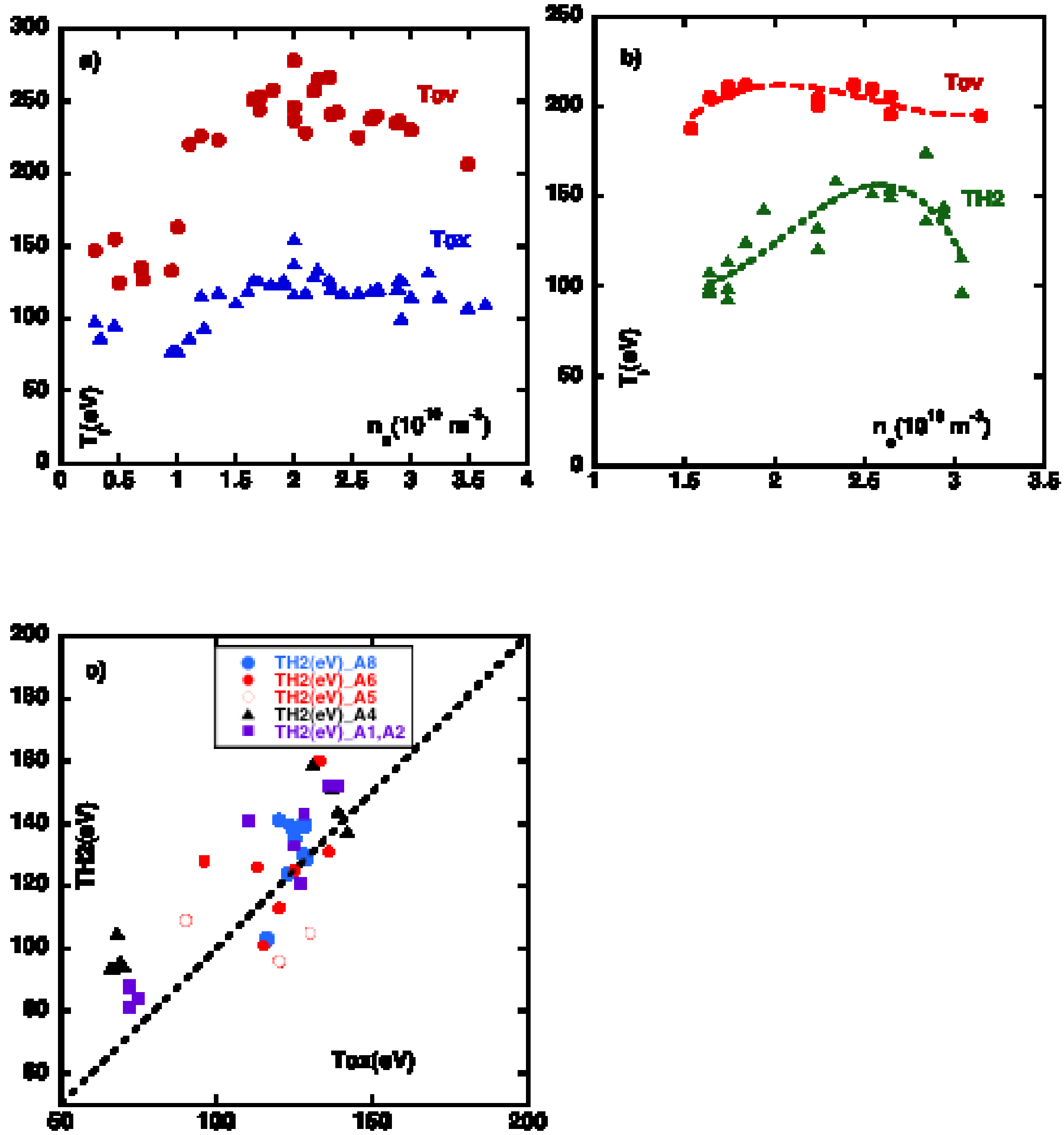
a)
Tov
Tox
T(eV)
$n_e(10^{19}\ m^{-3})$
b)
Tov
TH2
T(eV)
$n_e(10^{19}\ m^{-3})$
c)
TH2(eV)_A8
TH2(eV)_A6
TH2(eV)_A5
TH2(eV)_A4
TH2(eV)_A1,A2
TH2(eV)
Tox(eV)

Fig. 2

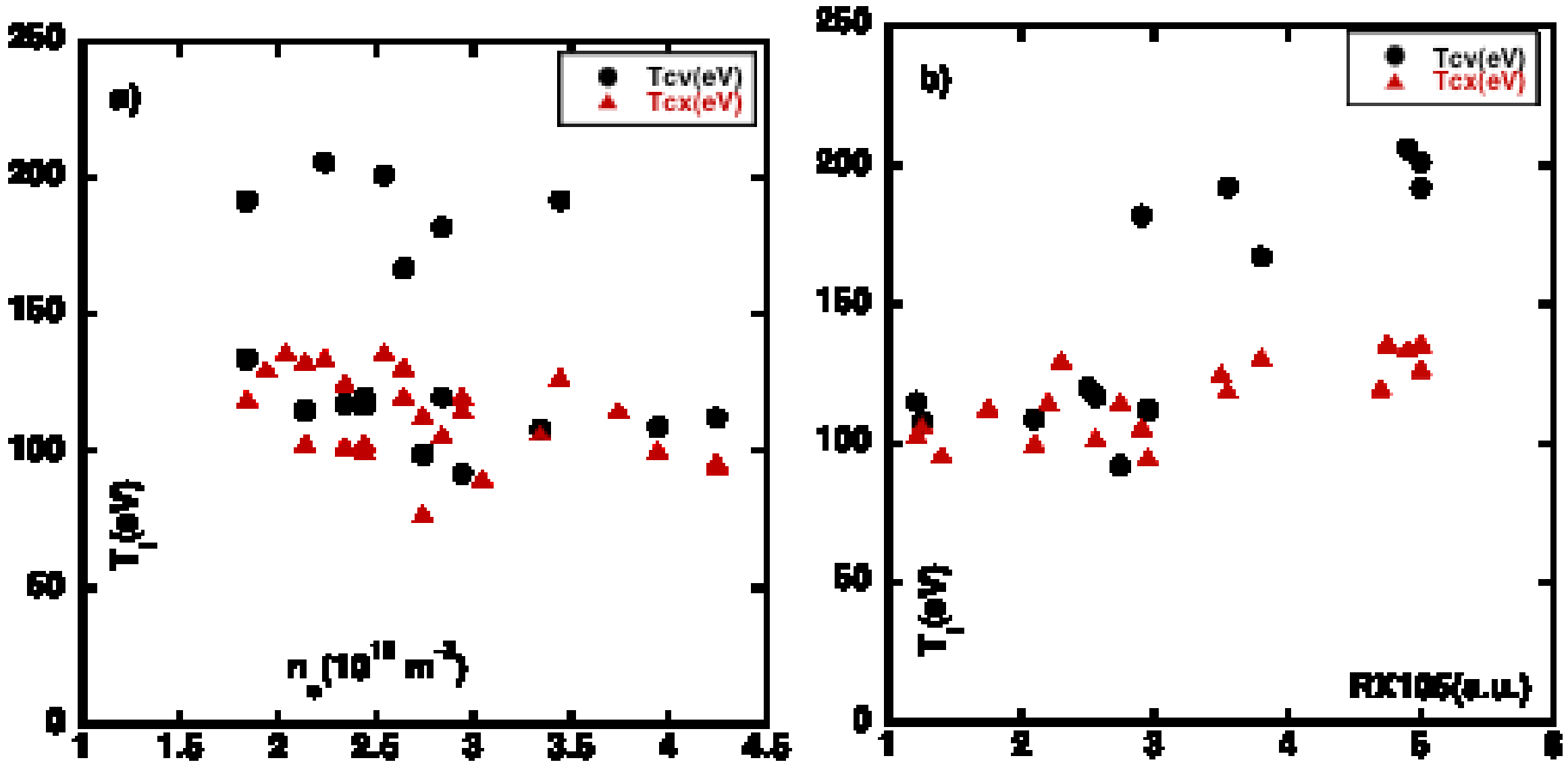
a)
Tcv(eV)
Tcx(eV)
T_i (eV)
n_e (10^19 m^-3)
250
200
150
100
50
0
1
1.5
2
2.5
3
3.5
4
4.5
b)
Tcv(eV)
Tcx(eV)
T_i (eV)
RX105(a.u.)
250
200
150
100
50
0
1
2
3
4
5
6

Fig. 3

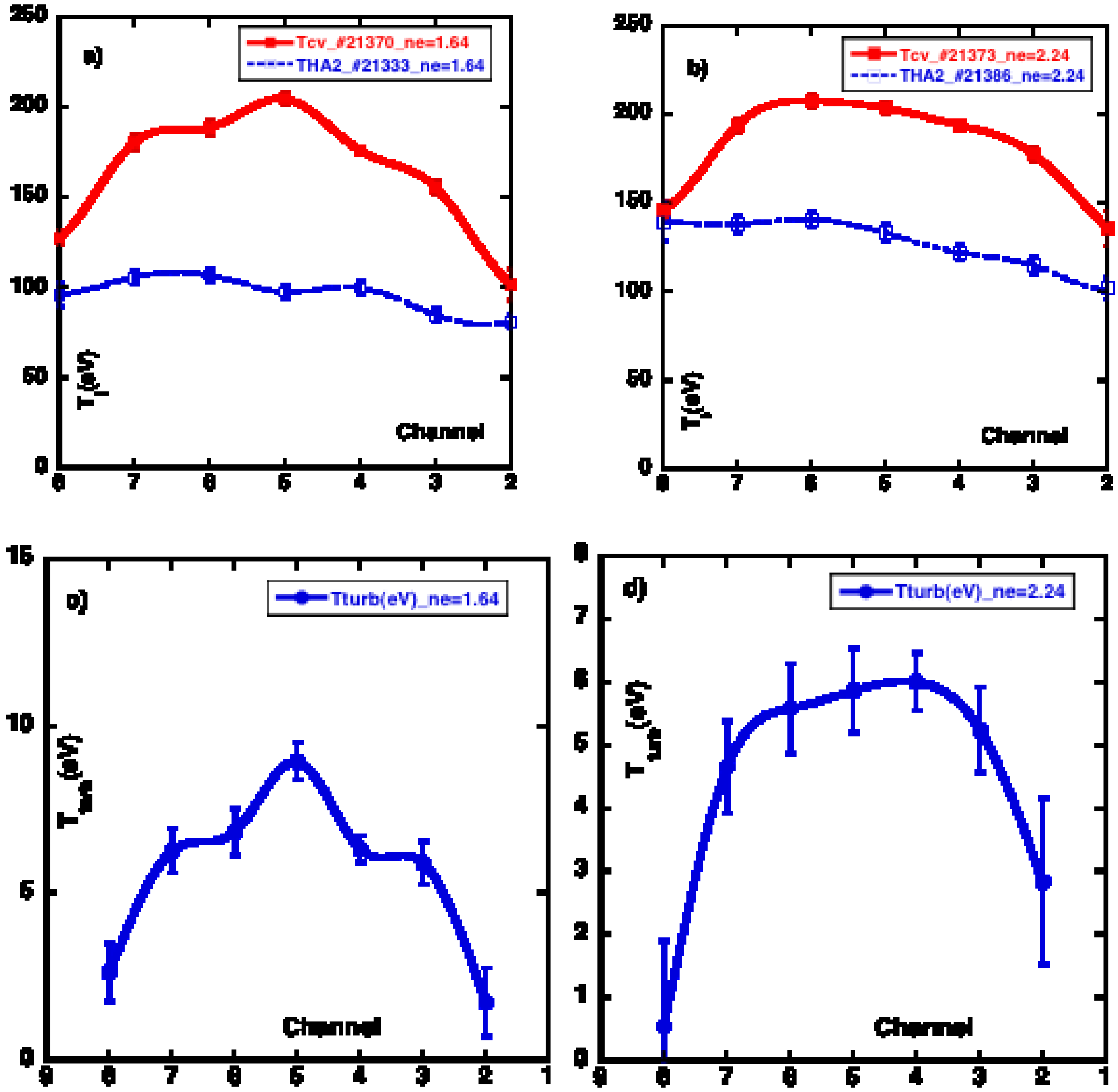

a)
Tcv_#21370_ne=1.64
THA2_#21333_ne=1.64
$T_i$(eV)
Channel
b)
Tcv_#21373_ne=2.24
THA2_#21386_ne=2.24
$T_i$(eV)
Channel
c)
Tturb(eV)_ne=1.64
$T_{turb}$(eV)
Channel
d)
Tturb(eV)_ne=2.24
$T_{turb}$(eV)
Channel

Fig. 4

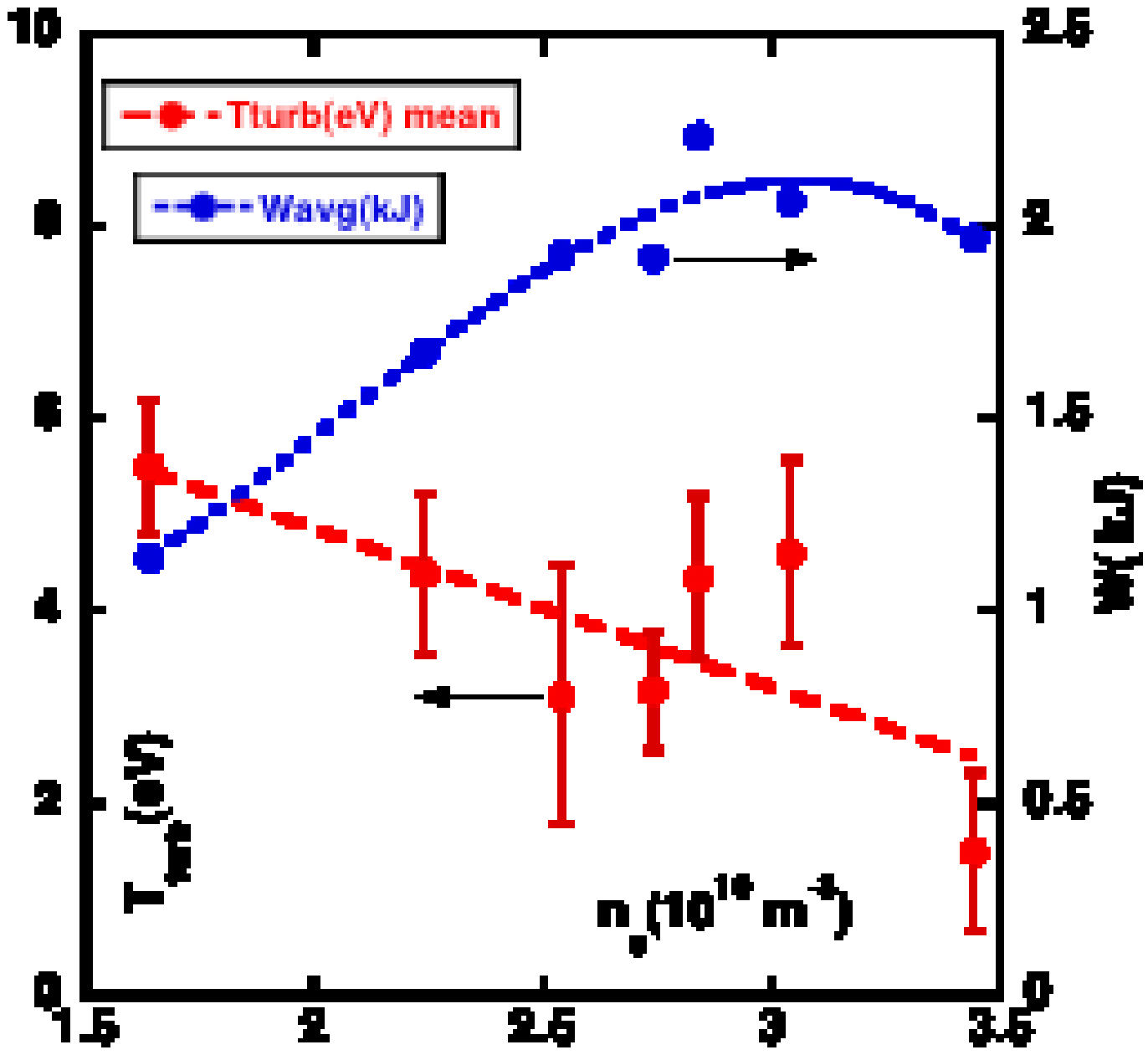

Tturb(eV) mean
Wavg(kJ)
T$_{turb}$ (eV)
W (kJ)
n$_e$(10$^{19}$ m$^{-3}$)
10
8
6
4
2
0
2.5
2
1.5
1
0.5
0
1.5
2
2.5
3
3.5

Fig. 5

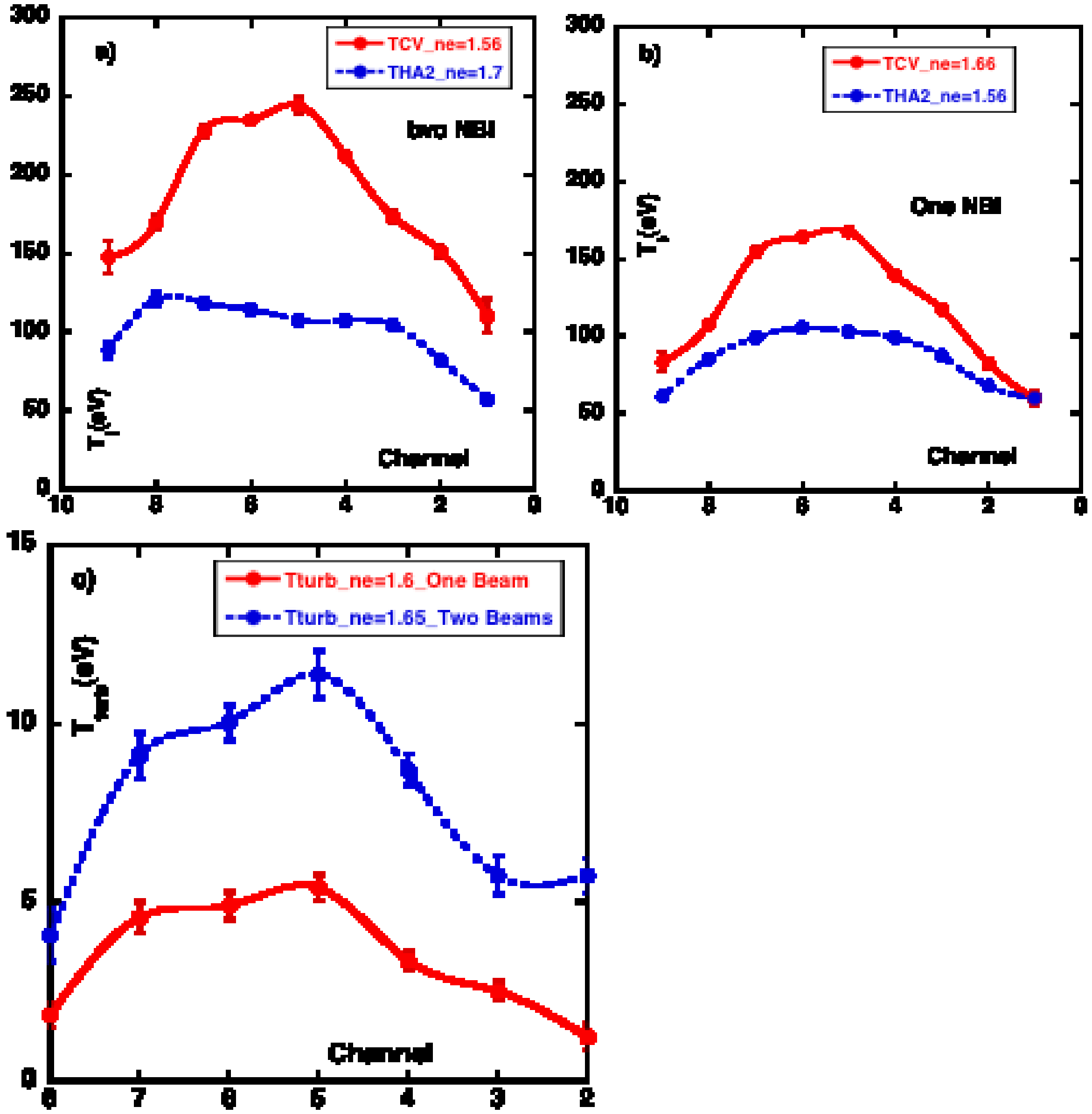
a)
TCV_ne=1.56
THA2_ne=1.7
two NBI
T (eV)
Channel
b)
TCV_ne=1.66
THA2_ne=1.56
One NBI
T (eV)
Channel
c)
Tturb_ne=1.6_One Beam
Tturb_ne=1.65_Two Beams
Channel